\documentclass{egpubl}
\usepackage{eg2025}
 
\ConferenceSubmission   
\usepackage[T1]{fontenc}
\usepackage{dfadobe}  

\usepackage{cite}  
\BibtexOrBiblatex
\electronicVersion
\PrintedOrElectronic
\ifpdf \usepackage[pdftex]{graphicx} \pdfcompresslevel=9
\else \usepackage[dvips]{graphicx} \fi

\usepackage{egweblnk} 
\usepackage{lineno}

\usepackage{mathtools}
\usepackage{amsfonts}
\usepackage{amssymb}
\usepackage{algorithm}
\usepackage{algpseudocode}
\usepackage{wrapfig}
\usepackage{cleveref}
\usepackage{xspace}
\usepackage{bm}
\usepackage[inline]{enumitem}
\usepackage{xcolor}
\usepackage{booktabs}
\usepackage{overpic}

\def\Reals{\mathbb{R}}

\newcommand{\mesh}[1]{\mathcal{#1}}
\def\M{\mesh{M}}
\def\N{\mesh{N}}
\def\S{\mesh{S}}

\newcommand{\BigO}[1]{\mathcal{O}\left(#1\right)}

\newcommand{\mymat}[1]{\mathbf{#1}}

\def\eg{\textit{e.g.}\xspace}
\def\ie{\textit{i.e.}\xspace}
\def\etal{\textit{et al.}\xspace}

\definecolor{BlueMATLAB}{HTML}{0072BD}
\definecolor{GreenMATLAB}{HTML}{77AC30}

\newcommand{\TableFirst}[1]{{\color{BlueMATLAB} \textbf{#1}}}
\newcommand{\TableSecond}[1]{{\color{GreenMATLAB} \textbf{#1}}}

\title[SShaDe: scalable shape deformation via local representations]%
      {SShaDe: scalable shape deformation via local representations}

\author[F. Maggioli \etal]
{\parbox{\textwidth}{\centering%
F. Maggioli$^{1}$\orcid{0000-0001-8008-8468},
D. Baieri$^{2}$\orcid{0000-0002-0704-5960},
Z. L\"{a}hner$^{3}$\orcid{0000-0003-0599-094X},
S. Melzi$^{1}$\orcid{0000-0003-2790-9591}
}
\\
{\parbox{\textwidth}{\centering%
$^1$University of Milano-Bicocca, Italy\\
$^2$Sapienza - University of Rome, Italy\\
$^3$University of Bonn, Germany
}
}
}

\begin{document}

\teaser{
 \includegraphics[width=\linewidth]{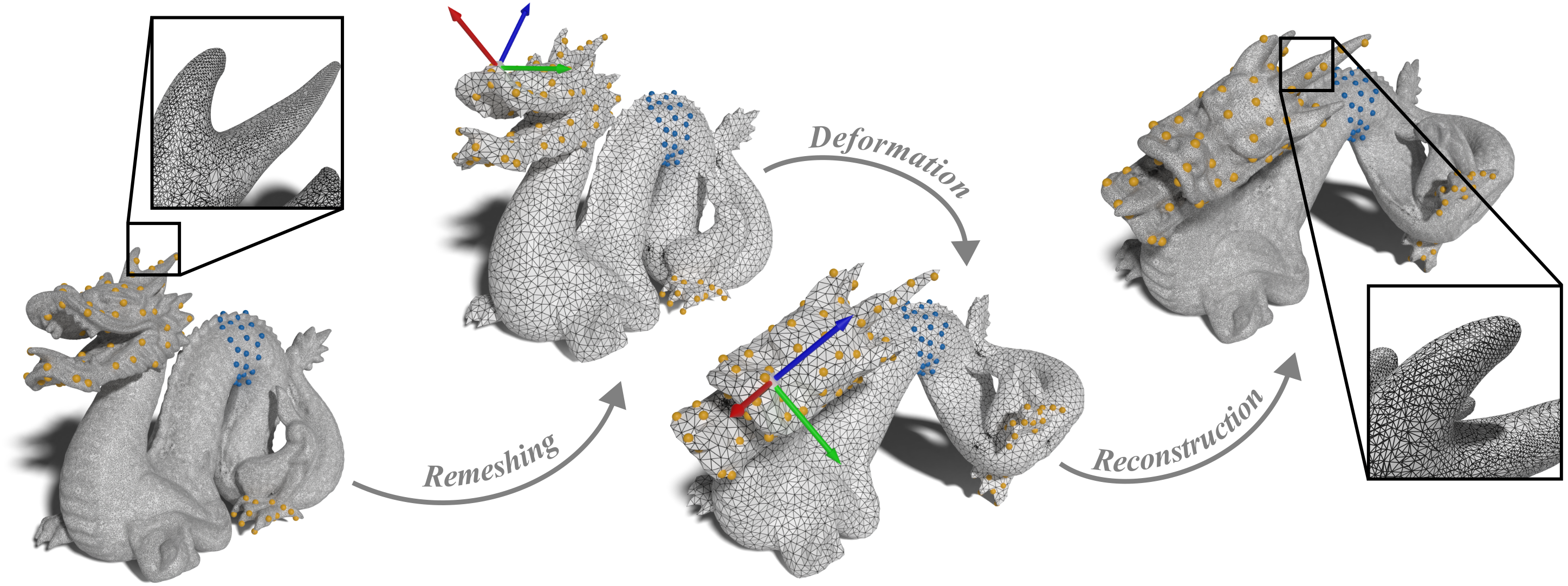}
 \centering
  \caption{A visual representation of our pipeline on handle-guided deformations. The input mesh (first) is remeshed to a lower resolution (second). The low resolution shape is then deformed (third), and the deformation of the original geometry (fourth) is reconstructed through a representation in a deformation-invariant local reference frame (second and third). Static handles are colored in blue, while dynamic handles are in yellow.}
\label{fig:teaser}
}

\maketitle
\begin{abstract}

With the increase in computational power for the available hardware, the demand for high-resolution data in computer graphics applications increases. Consequently, classical geometry processing techniques based on linear algebra solutions are starting to become obsolete. In this setting, we propose a novel approach for tackling mesh deformation tasks on high-resolution meshes. By reducing the input size with a fast remeshing technique and preserving a consistent representation of the original mesh with local reference frames, we provide a solution that is both scalable and robust in multiple applications, such as as-rigid-as-possible deformations, non-rigid isometric transformations, and pose transfer tasks. We extensively test our technique and compare it against state-of-the-art methods, proving that our approach can handle meshes with hundreds of thousands of vertices in tens of seconds while still achieving results comparable with the other solutions.


\end{abstract}  

\clearpage

\section{Introduction and related work}\label{sec:intro}
Deforming objects under the preservation of specific properties is one of the cornerstones of computer graphics, having countless applications in the field. 
Skeleton-based deformations~\cite{magnenat:1988:skeleton} are long-standing tools with widely applicability in surface animation. The simulation of deformable bodies is key in representing rubber-like materials~\cite{diziol:2011:incompressible}, cloth animations~\cite{bender:2013:cloth}, and natural processes and phenomena~\cite{maggioli:2023:plants}. Shape deformation has also been successfully used in shape matching tasks~\cite{eisenberger:2019:divergence}, and the interpolation of shapes has diverse useful applications in the field of geometry processing~\cite{solomon:2015:convolutional}. Additionally, recent advances in neural fields and Gaussian splatting drew some attention on the deformation~\cite{yang:2021:nfgp,baieri:2024:implicitarap}, editing~\cite{palandra:2024:gsedit,wu:2024:recent}, and simulation~\cite{zhong:2024:springgaus} with these types of representations.

One of the most famous and widely used deformation types is as-rigid-as-possible (ARAP), in which the target configuration should be reached while preserving the local rigid structure at all the surface points as much as possible~\cite{sorkine:2007:arap}. In the discrete setting, this translates to the simple constraint of preserving the edge lengths of the mesh. The powerful features that made ARAP deformations so popular are its simple formulation and the possibility of defining a small set of handle points, which can be used to drive the deformation. By combining the handle constraints and the local rigidity constraints, it is possible to obtain a visually plausible and natural looking global deformation.
Over the years many applications and extensions to this energy have been proposed; from a smoothness prior \cite{levi:2015:arapsre} to ARAP regularization terms \cite{huang2021arapreg} and incorporation into neural fields \cite{yuan2022nerf,baieri:2024:implicitarap}. 
However, optimizing for an optimal global ARAP deformation does not scale well to high-resolution shapes. Consequently, these types of algorithm are unsuitable for meshes with an high vertex count -- deforming meshes with more than 100k vertices with classical approaches can take several minutes on modern hardware.

A common workaround is to introduce a low-resolution reference graph on which the ARAP energy can be computed efficiently \cite{zorin1997interactive,habermann2020deepdynamic}. 
This leads to two challenges: 1) how to efficiently and accurately subsample the geometry while preserving the topology of the high-resolution mesh and 2) how to transfer the deformation computed on low-resolution vertices to the high-resolution geometry without introducing artifacts and errors. 
The task of obtaining a good sampling of the mesh is often a trade-off between efficiency and keeping correspondence information between the high- and low-resolution version. The front propagation method proposed by Peyr\'{e} \etal~\cite{peyre:2006:geodesic} proved to be very effective in computing a farthest point sampling of the surface. Maggioli \etal~\cite{maggioli:2023:rematching} refined this method by constructing a sampling of the surface that induces a low-resolution triangle mesh which is topologically equivalent to the input shape.
Obtaining a meaningful transfer of information from the low-resolution to the high-resolution is more challenging. For applications like avatar animation this is not a problem as a single deformation graph is sufficient \cite{habermann2020deepdynamic} but for general shapes retaining this information is not trivial \cite{jiang2020bijective}. 
However, this information is crucial for transferring the low-resolution information to the high-resolution vertices -- the high-resolution deformation should be obtained by mixing the deformations of the surrounding low-resolution vertices, but if not handled correctly, the interpolation might cause artifacts on boundaries and when the surface orientation changes.
Solutions for this problem in multi-resolution settings have been proposed for specific energies \cite{shi2006fast,yuan2022nerf} and data types \cite{zhou2005large,Yifan:NeuralCage:2020}, but they are not suitable for a more general setting where one has to deal with arbitrary meshes.
Another solution, more robust and general, has been proposed by Melzi \etal~\cite{melzi:2019:gframes} by using local reference frames at vertices. By using intrinsic quantities like surface normals and gradient vectors, the authors are able to define a local reference frame at each vertex that is invariant to isometric deformations of the shape.
More similar in spirit to our work is the contribution from Morsucci \etal~\cite{morsucci:2018:fastarap}, where the authors propose the construction of a low-resolution graph connectivity on which edge constraints can be easily defined and efficiently solved. The paper also proposes an approach for extending the deformation by interpolating the rigid transformations of the control points via a distance-based weighting. While this method proves to be both efficient and effective, it is only suitable for deformations that are defined via edge lengths, excluding a variety of possible applications.

With this paper, we presents a novel flexible approach to efficiently compute deformations through a low-resolution mesh and apply them to the original high-resolution shape. 
We leverage the fast remeshing of \cite{maggioli:2023:rematching} which provides a correspondence between the high- and low-resolution versions and perform an accurate reconstruction of geometric details by moving consistent local reference frames. 
Due to the knowledge of the correspondences this can be done very efficiently. 
Our experiments show a speed-up of 50x-100x in comparison to other ARAP implementations while keeping comparable metric scores. 
The method can also be used to tightly align isometrically deformed shapes and perform pose transfer between different classes. 

\paragraph*{Contributions. } Our contributions can be summed up as follows: 
\begin{itemize}
    \item A highly efficient framework for a broad range of shape deformation tasks that works by transferring detailed rigid geometry through local reference frames in a lower resolution. 
    \item An implementation that is 50 to 100 times faster than default ARAP deformation, which we applied to meshes with resolution up to 450k vertices with runtime performance of 10 seconds or less.
    \item Finally, we propose the first dataset for handle-guided deformations, automatically built from the well-known FAUST dataset by upsampling the original shapes and deriving motion handles from the ground truth correspondence.
\end{itemize}

\section{Background and notation}\label{sec:background}

\paragraph*{Triangular meshes.} We discretize a surface as a triangular mesh $\M = (V_{\M}, E_{\M}, T_{\M})$ embedded in $\Reals^3$, where:
\begin{itemize*}[label={}]
    \item $V_{\M} \subset \Reals^3$ is a set of vertices in the 3D space;
    \item $T_{\M} \subset V_{\M}^3$ is a set of triangular faces among vertices (invariant under even permutations);
    \item $E_{\M} \subset V_{\M}^2$ is a set of edges induced by the triangles (\ie, $(v_i, v_j, v_k) \in T_{\M} \implies (v_i, v_j) \in E_{\M}$).
\end{itemize*}

\paragraph*{Manifoldness.} Through this paper, we assume our meshes to be 2-manifold. This means that we don't allow a mesh $\M$ to have non-manifold edges (\ie, edges that are incident on three or more triangles), nor non-manifold vertices (\ie, vertices that are incident on two or more fans of triangles). We allow our meshes to have boundaries, that is to have some edges that are incident on a single triangle, but we assume them to be composed by a single connected component.

\paragraph*{Deformation.} When we refer to a deformation of a mesh $\M = (V_{\M}, E_{\M}, T_{\M})$, we indicate the creation of a second mesh $\M' = (V_{\M'}, E_{\M'}, T_{\M'})$ such that each vertex $v' \in V_{\M'}$ is obtained by altering the 3D position of a corresponding vertex $v \in V_{\M}$, but leaving unchanged the connectivity (\ie, $E_{\M'} = E_{\M}$ and $T_{\M'} = T_{\M}$). When the correspondence between vertices of a mesh $\M$ and its deformation $\M'$ is not implied in the text, we explicitly refer to it as a bijective function $\pi : V_{\M} \to V_{\M'}$ that preserves the connectivity isomorphism between the two meshes. Namely
\begin{gather}
    (v_i, v_j) \in E_{\M}
    \quad\iff\quad
    (\pi(v_i), \pi(v_j)) \in E_{\M'}\,,
    \\
    (v_i, v_j, v_k) \in T_{\M}
    \quad\iff\quad
    (\pi(v_i), \pi(v_j), \pi(v_k)) \in T_{\M'}\,.
\end{gather}

\paragraph*{Intrinsic quantities.} We use the term \emph{geodesic} as a shorthand for indicating a geodesic shortest path, and the term \emph{geodesic triangle} for indicating a convex region on a surface enclosed by the geodesics connecting three distinct surface points. Scalar functions over the surface are represented as functions $f : V_{\M} \to \Reals$ assuming real values at the vertices, while tangent vector fields $F : T_{\M} \to \Reals^3$ are defined on triangles. The surface normals are also defined at triangles, and for each triangle $t$ the orientation of its normal $n_t$ is defined by the orientation of its vertices using the right-hand rule. The gradient is an operator $\nabla : \mathcal{F}(\M, \Reals) \to \mathcal{F}(\M, \Reals^3)$ from the space $\mathcal{F}(\M, \Reals)$ of scalar functions over $\M$ to the space $\mathcal{F}(\M, \Reals^3)$ of vector fields over $\M$. The gradient $\nabla f(t)$ of a scalar function $f$ at a triangle $t$ returns the tangent vector at $t$ pointing towards the direction of the steepest increase in $f$.

\section{Method}\label{sec:method}

Our method achieves efficiency by optimizing a deformation on a low-resolution mesh but all information can be transferred accurately back to the original resolution. 
To that end, we first apply a topology-preserving remeshing to obtain a low-resolution version of the original shape while still preserving the overall geometry, see \Cref{sec:remeshing}. We then apply some deformation pipeline to the low-resolution mesh. Finally, we use a robust and conservative local reference frame (\Cref{sec:lrf}) to reconstruct the original geometry in the deformed pose, see \Cref{sec:refinement}.
An overview of our method can be found in \Cref{fig:teaser}.


\subsection{Remeshing}\label{sec:remeshing}

For the remeshing step, we require a procedure that
can efficiently reduce the vertex count by orders of magnitude, while still preserving the original topology and providing robustness guarantees about the preservation of the underlying geometry. 
Maggioli \etal~\cite{maggioli:2023:rematching} proposed a remeshing algorithm that obtains a farthest point sampling of the surface and constructs an intrinsic Delaunay triangulation from the dual Voronoi decomposition. The method associates each vertex $v$ and triangle $t$ of the remeshed shape to, respectively, a Voronoi region $R_v$ and a geodesic triangle $t_g$ on the high-resolution mesh. 
Given the original mesh $\M = (V_{\M}, E_{\M}, T_{\M})$ and its remesh $\N = (V_{\N}, E_{\N}, T_{\N})$, the method also builds an approximately bijective linear mapping $\mathbf{U} \in \Reals^{V_{\N} \times V_{\M}}$ for upsampling scalar functions over $\N$ to scalar functions over $\M$. Nonetheless, the map is built using a raw projection of the vertices $V_{\M}$ onto the nearest surface points of $\N$. This operation has a $\BigO{|V_{\M}| |T_{\N}|}$ cost, which can quickly become inefficient as the vertex count of $\M$ increases. 

\begin{wrapfigure}[10]{r}{0.35\columnwidth}
    \vspace{-1em}
    \hspace{-0.05\columnwidth}%
    \begin{overpic}[width=0.4\columnwidth]{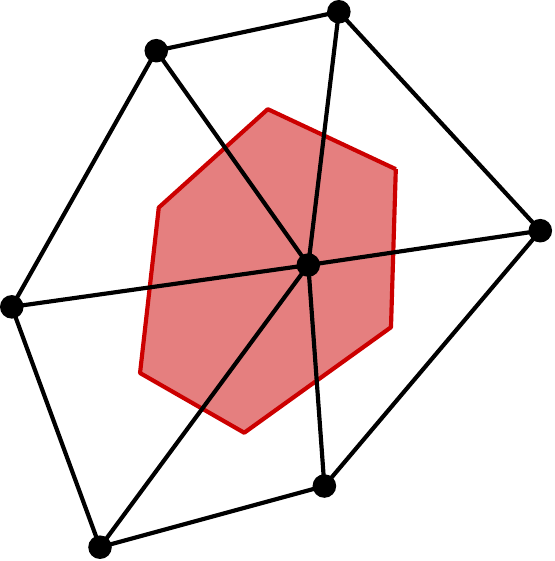}
    \put(60,57) {{\Large $v$}}
    \end{overpic}
\end{wrapfigure}
To solve this issue and further reduce the cost of the projection operation, we exploit the fact that each vertex $v \in V_{\N}$ is associated with a Voronoi region $R_v \subset V_{\M}$ onto $\M$, and that $R_v$ is a topological 2-disk. In this scenario, the original region $R_v$ must be represented by the dual Voronoi area of $v$ on $\N$, meaning that it is contained inside the triangle fan around $v$ (\ie, the dual Voronoi area of $v$ is the region of the mesh which is closer to $v$ than to any other vertex, as shown in red the inset figure). Thus, for each vertex $v \in V_{\N}$, we project the region $R_v$ onto the nearest surface points of the triangle fan around $v$. Statistically, the number of triangles incident on a vertex on a manifold mesh is always about 6-7, meaning that the projection operation for a Voronoi region has a $\BigO{|R_v|}$ cost. By summing across all the regions, we get
\begin{equation}
    \sum_{v \in V_{\N}}
    \BigO{|R_v|}
    =
    \BigO{|V_{\M}|}\,,
\end{equation}
achieving a significant improvement in the cost of the projection operation.

\begin{figure}[t]
    \centering
    \includegraphics[width=0.475\columnwidth]{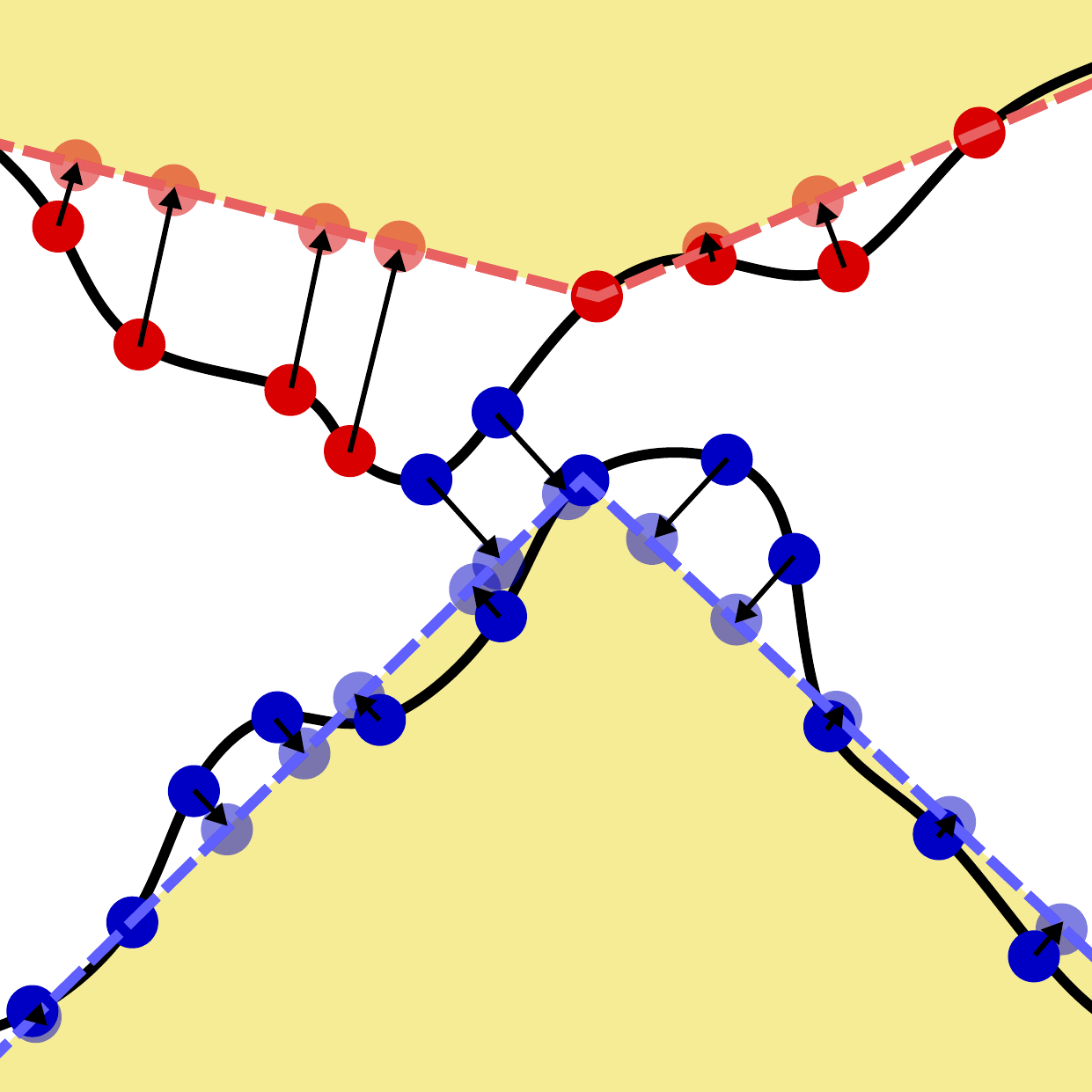}
    \hspace{\fill}
    \includegraphics[width=0.475\columnwidth]{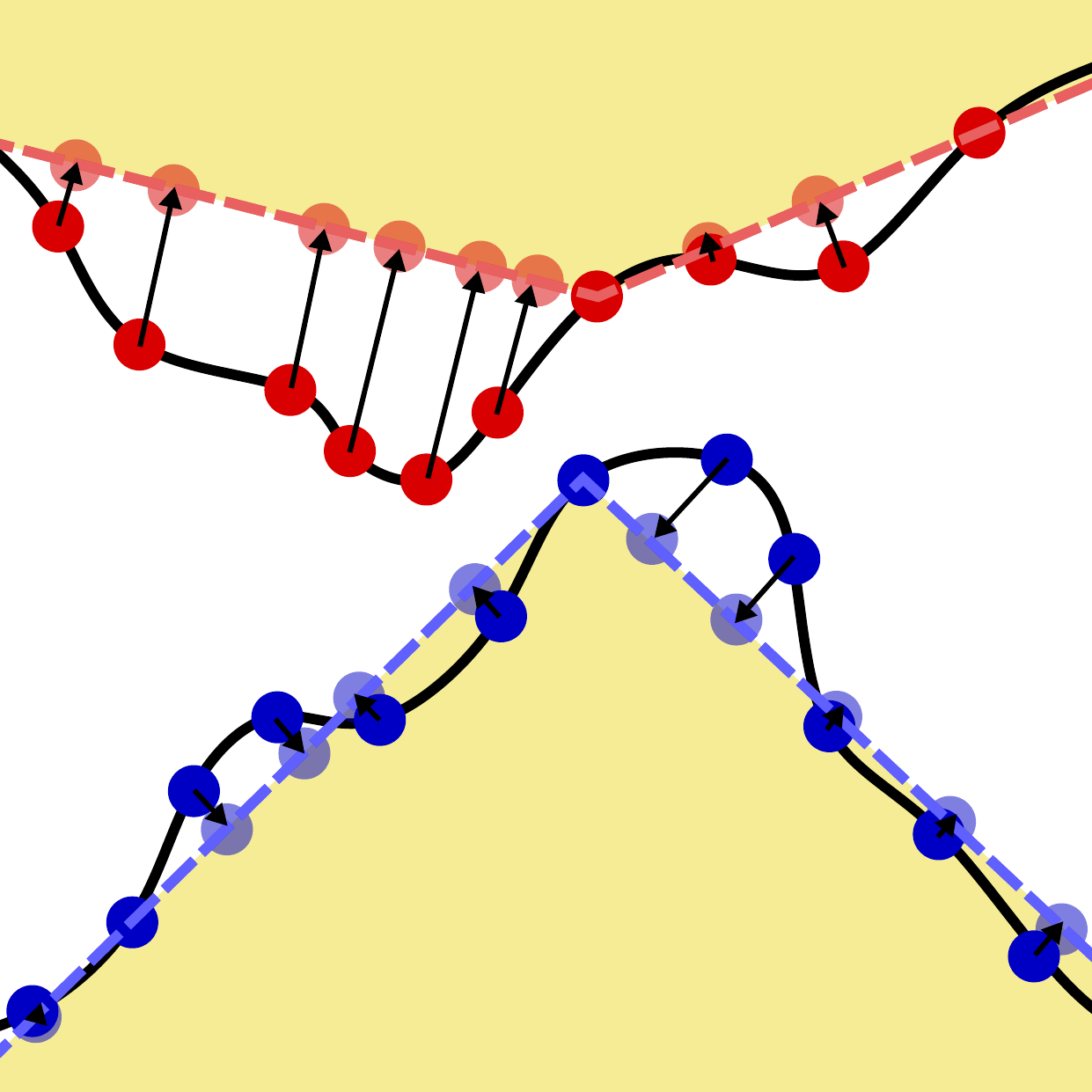}
    \caption{The vertices of two Voronoi regions on the high-resolution shape (solid) are projected onto the low-resolution representation (dashed, shaded in yellow) with the approach from \cite{maggioli:2023:rematching} (left) and our method (right). The color of the vertices (dark red or dark blue) reveals in which part of the low-resolution surface they are projected (respectively, light red or light blue).}
    \label{fig:projections}
\end{figure}

Furthermore,  this novel approach for computing the projection of $\M$ onto $\N$ has the additional desired property of mitigating artifacts and errors in the projection. As shown in the example from \Cref{fig:projections}, the original method proposed in \cite{maggioli:2023:rematching} does not account for regions of the surface that are geodesically far away, but close in the embedding space. Potentially, this situation results in an incorrect mapping of the vertices of $\M$ onto $\N$. On the other hand, with our approach we ensure that the vertices of each Voronoi 
 region of $\M$ are mapped into the correct region of $\N$.

\subsection{Local reference frame}\label{sec:lrf}

\begin{figure}[t]
    \centering
    \includegraphics[width=\columnwidth]{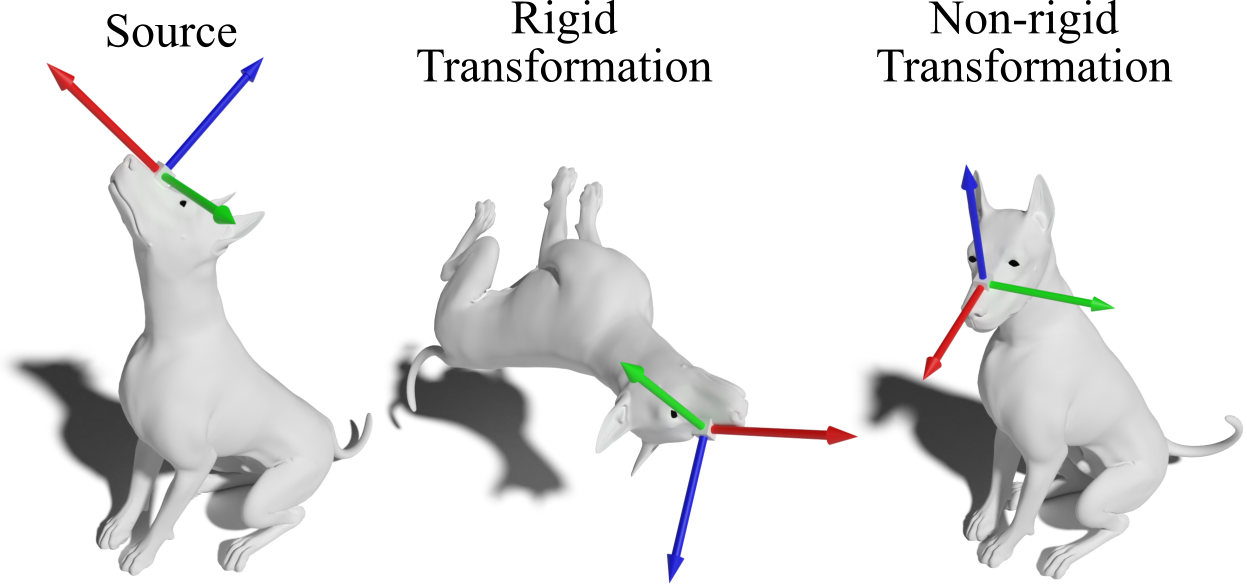}
    \caption{An example of local reference frame at a triangle (left) and how it is affected by a rigid transformation (center) and a non-rigid deformation (right) of the surface. The blue vector is always aligned with the surface normal, the red vector points towards the snout, and the green vector towards the left of the dog's head.
    }
    \label{fig:lrf-example}
\end{figure}

Given a mesh $\M = (V_{\M}, E_{\M}, T_{\M})$, the remeshing process produces a mesh $\N = (V_{\N}, E_{\N}, T_{\N})$, which represents $\M$ at a lower resolution, and a mapping $P : V_{\M} \to T_{\N}$ associating each vertex of $\M$ to the triangle of $\N$ it is projected onto. Since each triangle $t \in T_{\N}$ represents a geodesic triangle $t_g$ onto $\M$, by building a local reference frame at $t$, we can completely represent the geometry of $t_g$ in terms of local coordinates in $t$. Let $\N' = (V_{\N'}, E_{\N}, T_{\N})$ a connectivity-preserving deformation of the shape $\N$ (notice that edges and triangles are the same as $\N$), which may be obtained through an as-rigid-as-possible deformation, manual editing, or any other suitable deformation pipeline. If the local reference frame at triangle $t$ is consistent across deformations, we can use it to reconstruct the global coordinates of the geodesic triangle $t_g$ after the deformation, effectively deforming the original mesh $\M$ into a mesh $\M' = (V_{\M'}, E_{\M}, T_{\M})$. A visual representation of how local reference frames are deformed consistently through shapes is shown in~\Cref{fig:lrf-example}. As shown in the figure, the three basis vector always represent the same intrinsic directions and the same semantics across the shapes (\eg, the blue vector is always the surface normal and the red vector always points toward the snout).

Since we want our local reference frame to be consistent across deformations and variations of pose, we must build it using intrinsic measures and quantities. For this task, we adapt the approach proposed by Melzi \etal~\cite{melzi:2019:gframes}. In their paper, the authors propose to use the normal vectors and the normalized gradient of some pose-invariant scalar function $f : \N \to \Reals$ on the surface (\eg, the Laplacian eigenfunctions) to obtain a local reference frame at the vertices for shape correspondence tasks; the third vector can be easily obtained via a cross product between the other two. Differently from ~\cite{melzi:2019:gframes}, our setting provides us with a twofold advantage: 
\begin{itemize}
    \item we need our local reference frame to be defined over triangles rather than vertices;
    \item the correspondence between $\N$ and $\N’$ is already known.
\end{itemize}
Since triangles are the natural domain for normal and gradient vector, working on the mesh triangle allows us to reduce the number of operations to perform and improve numerical stability. 
Furthermore, given the correspondence between $\N$ and $\N'$ we can transfer any function accurately, thus enabling the usage of simple functions such as the mesh coordinates rather than pose-invariant functions, which are notoriously costly to compute and usually not robust on shapes at low-resolution.
Employing gradient vectors as basis vectors for local reference frames requires caution, as by the Poincar\'{e}-Hopf (``Hairy-Ball'') Theorem, not all surface topologies guarantee a non-null tangent vector field everywhere. In cases where we obtain tangent vectors with a very small magnitude at some triangles, we can introduce a second distinct function $f_2 : \N \to \Reals$ on the surface. For each triangle $t$ where $\nabla f(t)$ has a very small magnitude, we can replace the basis vector with with $\nabla f_2(t)$. 
However, as mentioned in \cite{melzi:2019:gframes}, in the discrete setting this situation occurs rarely, and we never experienced similar cases in our experiments.

After this process, for each triangle $t$ in the mesh $\N$, we obtain a matrix $\mymat{L}_t$ defined as
\begin{equation}
    \mymat{L}_t
    =
    \begin{pmatrix}
        g_t
        &
        w_t
        &
        n_t
    \end{pmatrix}\,,
\end{equation}
where $n_t$ is the normal vector at triangle $t$, $g_t$ is the gradient vector at $t$, and $w_t = g_t \times n_t$ is the cross product between the other two. By calling $b_t$ the barycenter of the triangle $t$, the representation of a vertex $v \in V_{\M}$ in the local reference frame of $t$ is given by
\begin{equation}
    v_t
    =
    \mymat{L}_t^{-1} \left( v - b_t \right)\,.
\end{equation}

Similarly, we can build a matrix $\mymat{L}'_t$ representing the local reference frame of the same triangle $t$, but in $\N'$. The reconstruction of the global coordinates for the deformed vertex $v' \in V_{\M'}$ can be obtained through
\begin{equation}
    v'
    =
    \mymat{L}'_t
    v_t
    +
    b'_t\,.
\end{equation}

\begin{algorithm}[t]
    \caption{Deformation transfer via a local reference frame.}
    \label{algo:lrf-transfer}
    \begin{algorithmic}[1]
        \Procedure{LRFT}{$\M$, $\N$, $\N'$, $f : V_{\N} \to \Reals$, $P : V_{\M} \to T_{\N}$}
            \For{$t \in T_{\N}$}
                \State $g_t, g'_t \gets$ gradient of $f$ at $t$ on $\N$ and $\N'$, respectively
                \State $n_t, n'_t \gets$ normal at $t$ on $\N$ and $\N'$, respectively
                \State $w_t \gets g_t \times n_t, \quad w'_t \gets g'_t \times n'_t$
                \State $b_t, b'_t \gets$ barycenter of $t$ in $\N$ and $\N'$, respectively
                \State $\mymat{L}_t \gets \left( g_t\ \ w_t\ \ n_t \right), \quad \mymat{L}'_t \gets \left( g'_t\ \ w'_t\ \ n'_t \right)$
            \EndFor
            \State $V_{\M'} \gets \varnothing$
            \For{$v \in V_{\M}$}
            \State $t \gets P(v)$
                \State $v_t \gets \mymat{L}_t^{-1} \left( v - b_t \right)$
                \State $v' \gets \mymat{L}'_t v_t + b'_t$
                \State $V_{\M'} \gets [ V_{\M'},\ v']$
            \EndFor
            \State \Return $\M' = (V_{\M'}, E_{\M}, T_{\M})$
        \EndProcedure
    \end{algorithmic}
\end{algorithm}

The entire procedure is summarized in \Cref{algo:lrf-transfer}, which accepts in input the high-resolution shape $\M$, its remesh $\N$, the deformed remesh $\N'$, a scalar function $f$ over $\N$, and the projection map $P$ of the vertices of $\M$ onto the triangle of $\N$. The first \textbf{for} loop (lines 2-8) computes the local reference frames of all the triangles in both the remeshed shape $\N$ and its deformation $\N'$. These local reference frames are used in the second \textbf{for} loop (lines 10-15) for representing the vertices of $\M$ in local coordinates and deforming them according to $\N'$.

\subsection{Refinement}\label{sec:refinement}

\begin{figure}[t]
    \centering
    \includegraphics[width=\columnwidth]{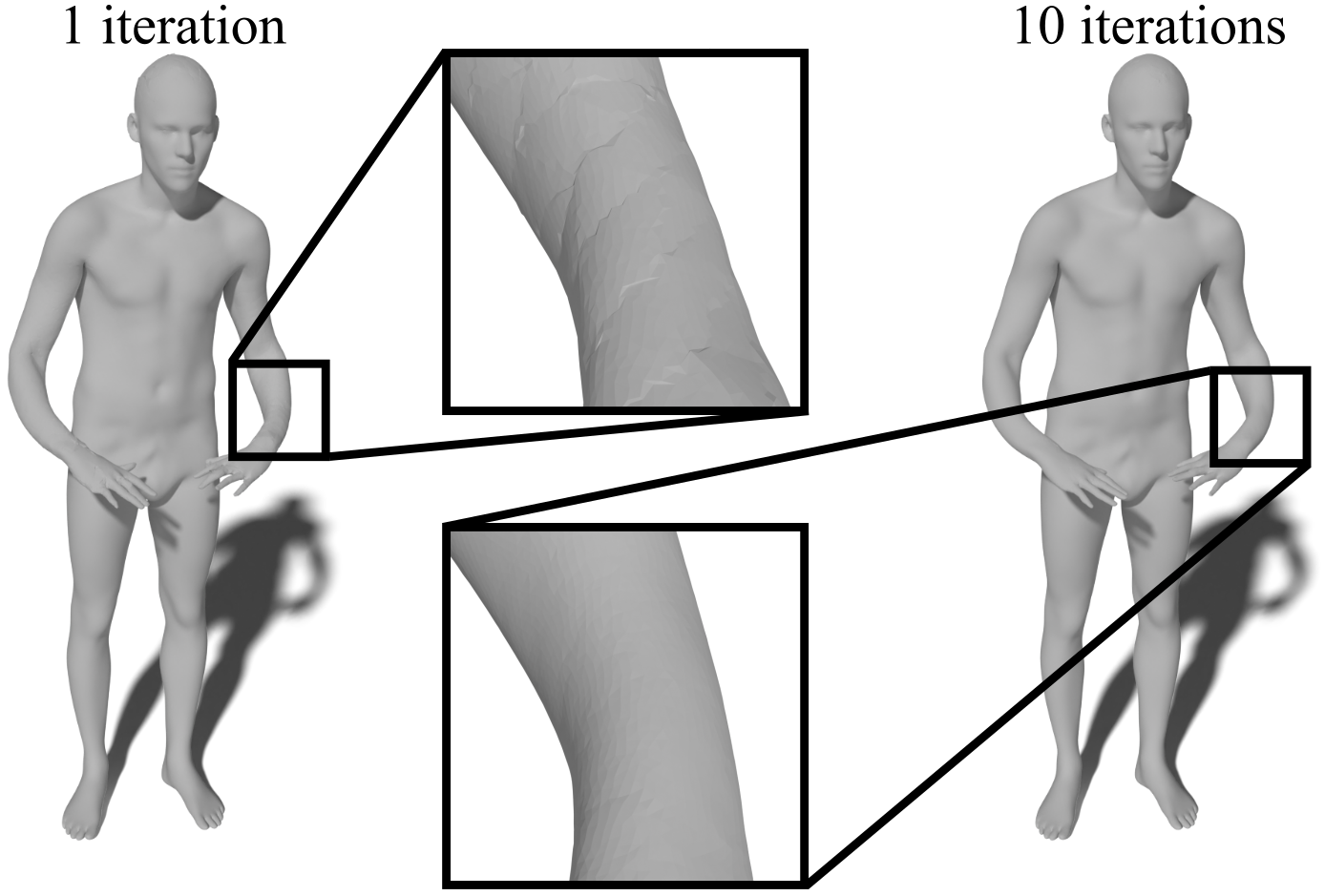}
    \caption{Comparison between the direct application of our approach (left) and the averaging of multiple iterations (right). The closeups on the arms show the difference in smoothness between the two approaches.}
    \label{fig:average-deformation}
\end{figure}

The approach we described so far has a major drawback due to the linearity of the local reference frames. 
Indeed, representing a local reference frame as an affine 3D transformation allows for fast computation, but it also limits the representational power of the deformation. Each geodesic triangle $t_g$ on $\M$ associated with a triangle $t \in T_{\N}$ is deformed according to the same linear transformation induced by the local reference frame. As a result, the deformed mesh at high-resolution presents visual artifacts and lines along the edges of the geodesic triangles, as can be seen in \Cref{fig:average-deformation}. In order to mitigate this effect, we apply multiple iterations of our process using different samplings for computing the low-resolution mesh, resulting in different intrinsic triangulations of $\M$. Each of these iterations produces a slightly different version of the final mesh $\M’$, mostly differing only for the placement of the artifact lines. By averaging all these versions of $\M’$, we obtain a final result where the artifacts are smoothed out, but which still preserves all the details of the original surface. While in our experiments we notice that 5-10 iterations are generally enough, we acknowledge that estimating an optimal number of iterations could be worth of future investigations. We stress that, as each individual iteration is independent, the process can be easily implemented in a multi-threaded environment, averaging all the results at the end and achieving an almost perfect parallelism.

\section{Results}\label{sec:results}

\begin{figure*}[t]
    \centering
    \includegraphics[width=\linewidth]{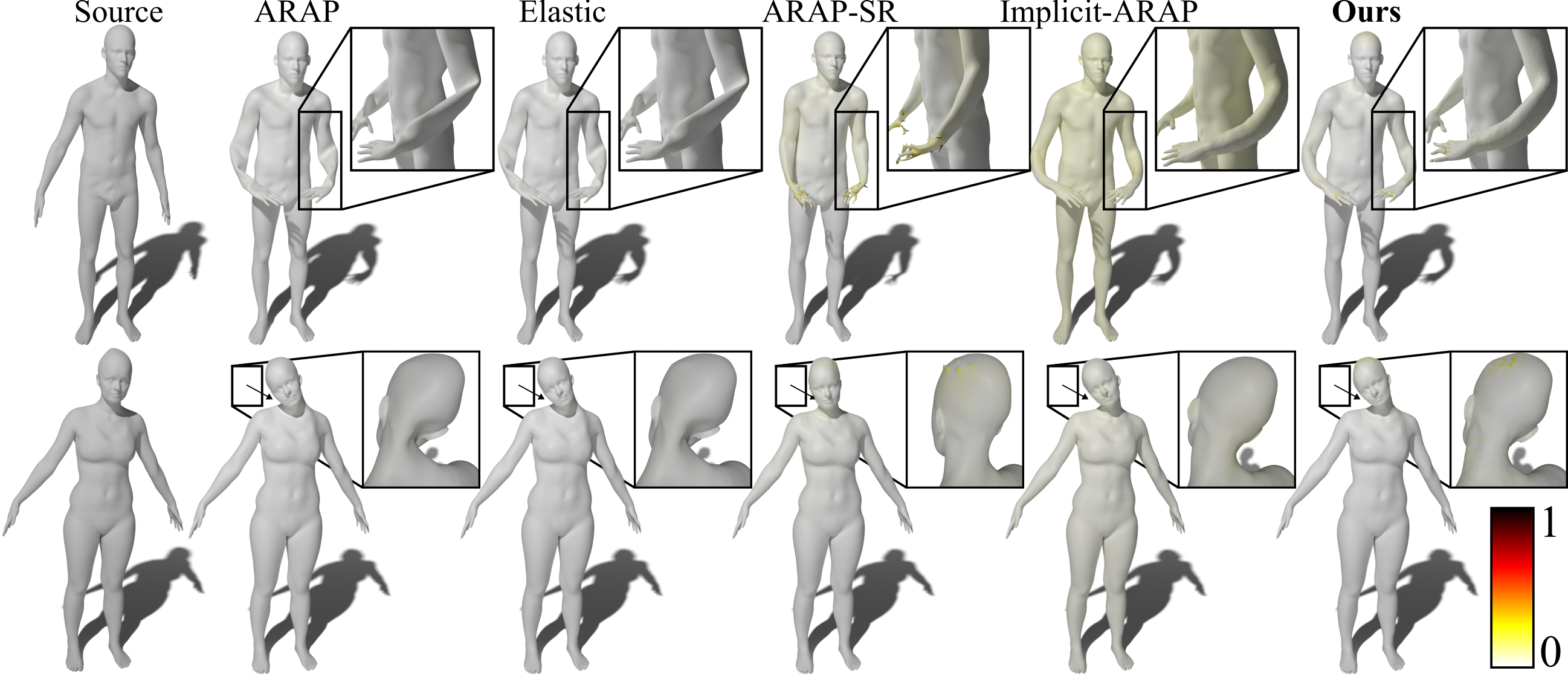}
    \caption{Comparison of our solution with the baseline approaches on sample deformations from the DeFAUST dataset. The close-ups show the deformation artifacts, and the edge error as in \Cref{eq:edge-error} is shown as a hot colormap.}
    \label{fig:defaust-volume}
\end{figure*}

We validate the performance of our method by testing its performance on handle-guided deformation tasks against state-of-the-art solutions. We also evaluate the pipeline on different types of deformations, to prove its applicability in different contexts. Finally, we show that our approach can be exploited for pose transfer tasks.

Our method is implemented in C++, using as backends the Eigen library for linear algebra routines~\cite{eigenweb} and the libigl library for basic geometric operations~\cite{libigl}. All the experiments are performed on a machine equipped with a CPU Intel i7-10700K and 32GB of RAM. 

\subsection{As-rigid-as-possible deformations}\label{sec:results:arap}

As-rigid-as-possible (ARAP) deformations represent a core task in the field of geometry processing but most existing solutions are known to suffer from poor scalability to high-resolution meshes. 
To evaluate the performance of our pipeline in this setting, we compare it against the original ARAP implementation (ARAP)~\cite{sorkine:2007:arap}, the smooth rotations variant (ARAP-SR)~\cite{levi:2015:arapsre}, and the spokes and rims method (Elastic)~\cite{chao:2010:arapsar}. 
The latter is also the method we use for deforming our low-resolution shapes, as it is the method that gives the best trade-off between quality and computational cost. 
Additionally, we consider the recent work from Baieri \etal (Implicit-ARAP)~\cite{baieri:2024:implicitarap}, as it provides an efficient and more scalable solutions to previous approaches while still achieving comparable quality. Finally, we did not consider the work from Morsucci \etal~\cite{morsucci:2018:fastarap} for the evaluation, as there is no publicly available implementation.
For our experiments, we use some of the user-defined deformations used in \cite{baieri:2024:implicitarap}, which are based on models from the Stanford's 3D Scanning Repository (S3D)~\cite{stanford3d}.
Moreover, we test all the competitors on the DeFAUST dataset (see \Cref{sec:defaust} for an in-depth presentation of the dataset), a novel dataset that we propose to validate accuracy and efficiency in this task on high-resolution meshes. This evaluation enables us to provide a quantitative comparison that was completely missing on this task and thus is an additional concrete contribution.

For evaluating the results, we follow the approach from \cite{baieri:2024:implicitarap}. Given the original surface $\M$ and the deformed surface $\M'$, we consider the relative surface error $E_{\mathrm{area}}$ and volume error $E_{\mathrm{vol}}$
\begin{equation}\label{eq:surf-vol-err}
    E_{\mathrm{area}}
    =
    \frac{
        \left\lvert A_{\M} - A_{\M'} \right\rvert
    }{
        A_{\M}
    }
    \,,
    \quad
    E_{\mathrm{vol}}
    =
    \frac{
        \left\lvert V_{\M} - V_{\M'} \right\rvert
    }{
        V_{\M}
    }\,,
\end{equation}
where $A_{\mesh{S}}$ and $V_{\mesh{S}}$ are, respectively, the surface area and the volume of a watertight surface $\mesh{S}$. We also evaluate the distortion induced on the input geometry by considering the edge length error and the face angle error. Namely, the edge length error is the average difference in length between the same edge before and after the deformation, normalized by the maximum edge length. This can be computed as
\begin{equation}\label{eq:edge-error}
    E_{\mathrm{edge}}
    =
    \frac{1}{| E_{\M} |}
    \sum_{(u, v) \in E_{\M}}
    \frac{\left\lvert \| u - v \| - \| \pi(u) - \pi(v) \| \right\rvert}{\max_{e \in E_{\M}} \|e\|}\,,
\end{equation}
where $\pi : V_{\M} \to V_{\M'}$ is the point-to-point correspondence between the vertices of the original mesh $\M$ and its deformation $\M'$. Similarly, the face angle error is the average difference between the same internal angles of the triangles before and after the deformation, which is given by
\begin{equation}\label{eq:angle-error}
    E_{\mathrm{angle}}
    =
    \frac{1}{3 | T_{\M} |}
    \sum_{t \in T_{\M}}\sum_{v \in t}
    \frac{\left\lvert \angle(v, t) - \angle(\pi(v), t) \right\rvert}{\angle(v, t)}
    \,,
\end{equation}
where $\angle(v, t)$ is the internal angle of $t$ at vertex $v$.

\begin{table}[t]
    \centering
    {\footnotesize%
    \begin{tabular}{c | c  c  c  c | c}
    Method & Volume & Area & EL & FA & Time \\
    \hline
    ARAP                                & 
    12.41\%                             & 
    \TableFirst{0.45\%}                 & 
    \TableFirst{0.37\%}                 & 
    \TableFirst{0.517\textdegree}       & 
    7m:15s                              
    \\
    Elastic                             & 
    12.25\%                             & 
    \TableFirst{0.45\%}                 & 
    \TableSecond{0.38\%}                & 
    \TableSecond{0.523\textdegree}      & 
    8m:27s 
    \\
    ARAP-SR                             & 
    11.69\%                             & 
    5.36\%                              & 
    3.39\%                              & 
    5.628\textdegree                    & 
    \TableSecond{7m:11s} 
    \\
    Implicit-ARAP                       & 
    \TableFirst{0.99\%}                 & 
    0.82\%                              & 
    1.06\%                              & 
    1.517\textdegree                    & 
    8m:22s 
    \\
    \textbf{Ours}                       & 
    \TableSecond{2.28\%}                & 
    \TableSecond{0.59\%}                & 
    1.28\%                              & 
    2.148\textdegree                    & 
    \TableFirst{0m:8s}
\end{tabular}%
    }
    \caption{Average errors and execution times for our method and the considered baseline approaches on the DeFAUST dataset. The angle error (FA) is expressed in degrees, while the edge length error (EL), the surface error, and the volume error are in percentage. For each column, the best and second best results are highlighted in light blue and light green, respectively.}
    \label{tab:defaust-results}
    \vspace{-1em}
\end{table}

\begin{figure*}[t]
    \centering
    \includegraphics[width=\linewidth]{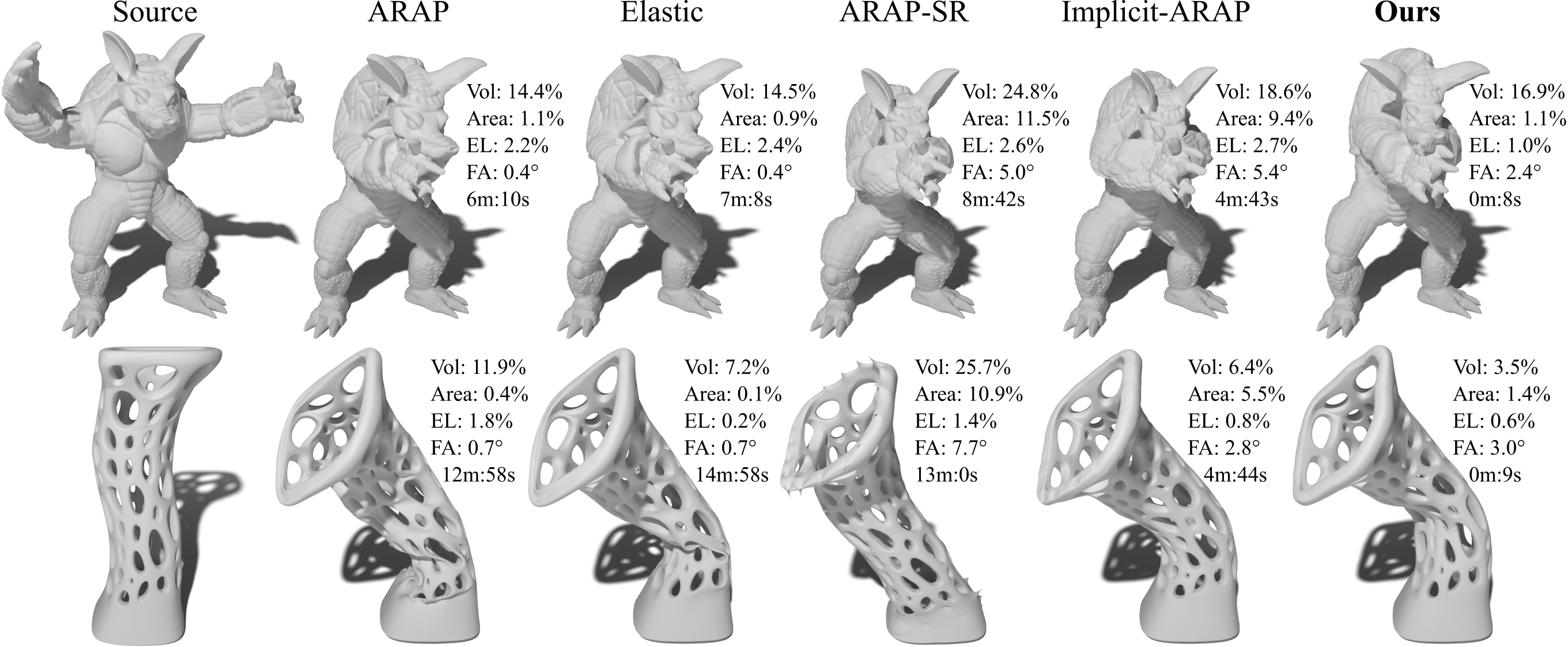}
    \caption{Comparison of our solution with the baseline approaches on some user-defined deformations from \cite{baieri:2024:implicitarap}.}
    \label{fig:arap-examples}
\end{figure*}

The results of our experiments on the DeFAUST dataset are summarized in \Cref{tab:defaust-results} where we show the error metrics and the runtime averaged across all the 40 high-resolution examples in the dataset. 
The experiments are run until convergence with a cutoff of 1000 iterations. For our method, we remesh the original surface to about 6000 vertices, averaging the result across 5 parallel iterations (as discussed in \Cref{sec:refinement}). 
The configuration parameters of Implicit-ARAP are set to the values suggested by the authors in their paper. 
The experiments show that our approach can achieve comparable results to the other baseline approaches in terms of surface error, edge length error, and face angle error. However, with our method we can compute the deformation at a fraction of the computational cost, obtaining a 50x-60x speedup. 
Notably, both Implicit-ARAP and our solution achieve significantly better results in terms of volume error. This behavior can be better appreciated in the examples from \Cref{fig:defaust-volume}, which also shows the ability of our method to keep the per-edge error low. 
We speculate that this outcome on volume preservation comes from an intrinsic issue of ARAP and its variants, as the local rigidity constraints only translate to useful volume preservation in a low resolution setting. 
Clearly, this is not an issue for continuous implicit methods such as Implicit-ARAP, while our method may be seen specifically as a way to account for this issue, since we perform the deformation in a low-resolution representation of the original input. In support to this argument, we highlight that the volume errors for the baselines in the same exact set of experiments on the low-resolution FAUST shapes are much more competitive (ARAP: 1.28\%, Elastic: 1.29\%, ARAP-SR: 1.21\%). 

Lastly, the examples in \Cref{fig:arap-examples} depict a comparison between our method and the other baselines on some user-defined deformations from the original Implicit-ARAP paper. As shown in the figure, with our approach we are able to obtain results that are more visually plausible with few seconds of computation, instead of several minutes.

\subsection{Non-rigid isometric deformations}\label{sec:results:isometric}

Our pipeline is not bound to ARAP deformations only: as long as the deformation does not change the number of vertices and their connectivity, our method can be applied to other types of deformations as well. In this regard, we design a set of experiments to test the performance of our pipeline when tackling arbitrary non-rigid isometric deformations. 
We use the TOSCA dataset~\cite{bronstein:2008:tosca} which is composed by synthetic meshes representing 9 human and animal subjects in several different poses, including a neutral rest pose. By using the provided correspondence and starting from the rest pose, we reconstruct all the other poses from a low-resolution deformation. Given the rest pose $\M = (V_{\M}, E_{\M}, T_{\M})$ of a subject and another arbitrary pose $\M' = (V_{\M'}, E_{\M'}, T_{\M'})$ of the same subject, we apply the following procedure:

\begin{enumerate}[noitemsep,topsep=0pt]
    \item we remesh $\M$ to a low-resolution version $\N = (V_{\N}, E_{\N}, T_{\N})$;
    \item since $V_{\N}$ is a subset of $V_{\M}$, we use the correspondence provided by the TOSCA dataset to obtain the corresponding vertices $V_{\N'}$ in the target pose;
    \item we infer the remesh of the target pose as $\N' = (V_{\N'}, E_{\N}, T_{\N})$ using the same connectivity as $\N$;
    \item using the representation of $\M$ in the local reference frame of the triangles of $\N$, we transfer the local coordinates onto $\N'$ and reconstruct the deformed mesh $\M'$.
\end{enumerate}

\begin{figure}[t]
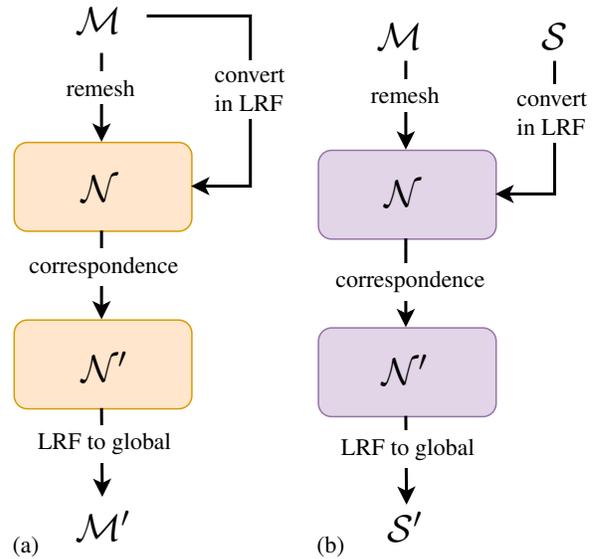

    \centering
    \hspace{\fill}%
    \input{imgs/tosca-deform-pipeline/overpic}%
    \hspace{\fill}%
    \input{imgs/pose-transfer-pipeline/overpic}
    \hspace{\fill}%
    \caption{The pipelines used for transferring the non-isometric deformations from the TOSCA dataset (a) and for the pose transfer tasks on the manually edited meshes (b).}
    \label{fig:diagrams}
    \vspace{-2em}
\end{figure}

\begin{figure}[t]
    \centering
    \includegraphics[width=\columnwidth]{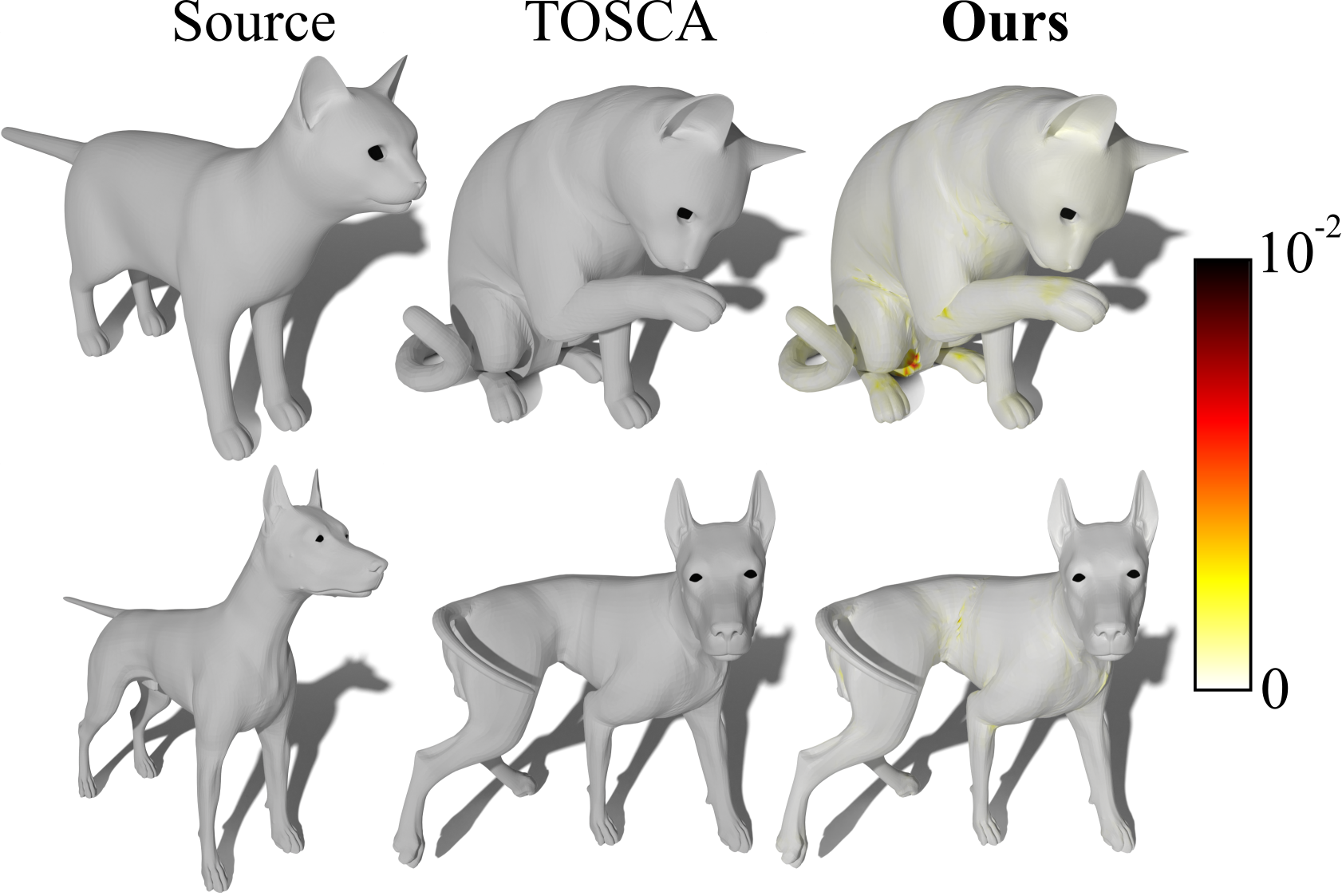}
    \caption{From left to right: a mesh from the TOSCA dataset representing a subject in its rest pose; a mesh from the TOSCA dataset representing the same subject in another pose; the mesh obtained by deforming the rest pose with our pipeline. The Euclidean distance between the deformed mesh and the nearest surface point of the TOSCA mesh is shown as a function on the surface.}
    \label{fig:tosca-isometric}
\end{figure}

The approach is summarized by the diagram in \Cref{fig:diagrams}a.
In \Cref{fig:tosca-isometric} we provide visual comparisons between the original meshes from the TOSCA dataset and those reconstructed with our approach. Here we remesh the surfaces to about 6000 vertices and average the result across 10 deformations. We apply this pipeline to the entire TOSCA dataset, comparing the original TOSCA meshes and our reconstruction. We evaluate the results by computing the Hausdorff distance $d_H$ and the Chamfer distance $d_C$, which are defined as
\begin{gather}
    d_H(\mesh{X}, \mesh{Y})
    =
    \max\left(
    \max_{x \in V_{\mesh{X}}} d(x, \mesh{Y}),\ 
    \max_{y \in V_{\mesh{Y}}} d(y, \mesh{X})
    \right)\,,
    \\
    d_C(\mesh{X}, \mesh{Y})
    =
    \frac{1}{|V_{\mesh{X}}|}
    \sum_{x \in V_{\mesh{X}}} d^2(x, \mesh{Y})
    +
    \frac{1}{|V_{\mesh{Y}}|}
    \sum_{y \in V_{\mesh{Y}}} d^2(y, \mesh{X})\,,
\end{gather}
\noindent
where $d(v, \mesh{S})$ represents the Euclidean distance between the point $v$ and the surface $\mesh{S}$. For a meaningful comparison, we center all the shapes at the origin of the axes and rescale them to fit inside the unit sphere. Furthermore, we exclude from the experiments the \texttt{wolf} shape for its low vertex count (less than 5000 vertices) and the \texttt{gorilla} shape for its non-manifoldness and high number of connected components. By averaging the results across the entire dataset, we obtain a Hausdorff distance $d_H = 5.66 \cdot 10^{-3}$ and a Chamfer distance $d_C = 3.76 \cdot 10^{-8}$.
\Cref{fig:tosca-isometric} shows how tight the alignment is.

\subsection{Pose transfer}\label{sec:results:posetransfer}

Another possible use-case scenario for our method is the transfer of the pose of a shape into the one of another shape. 
To demonstrate the effectiveness of our method for this task, we manually edit the rest pose of the \texttt{centaur} mesh and the \texttt{cat} mesh from the TOSCA dataset to obtain two new meshes that are semantically similar to the original but present different features. 
Notice that the new meshes are in one-to-one correspondence with the original shapes, share the same connectivity, and are aligned in 3D space.
Given the rest pose $\M$ from the TOSCA dataset, the edited mesh $\S$ in the rest pose, and a mesh $\M'$ representing the subject $\M$ in a different pose, we want to compute a deformation $\S'$ of $\S$ representing the edited subject in the pose provided by $\M'$. We remesh the original mesh $\M$ into a low-resolution shape $\N$, and follow the process summarized by \Cref{fig:diagrams}b. The procedure is similar to the one we describe in \Cref{sec:results:isometric} but instead of computing the local representation of $\M$ onto $\N$, we use the one-to-one correspondence between $\M$ and $\S$ to compute the local coordinates of the vertices of $\S$ onto the triangles of $\N$. Then, we reconstruct the mesh $\S'$ representing the edited mesh $\S$ into the target pose provided by $\M'$.

\begin{figure}[t]
    \centering
    \includegraphics[width=\columnwidth]{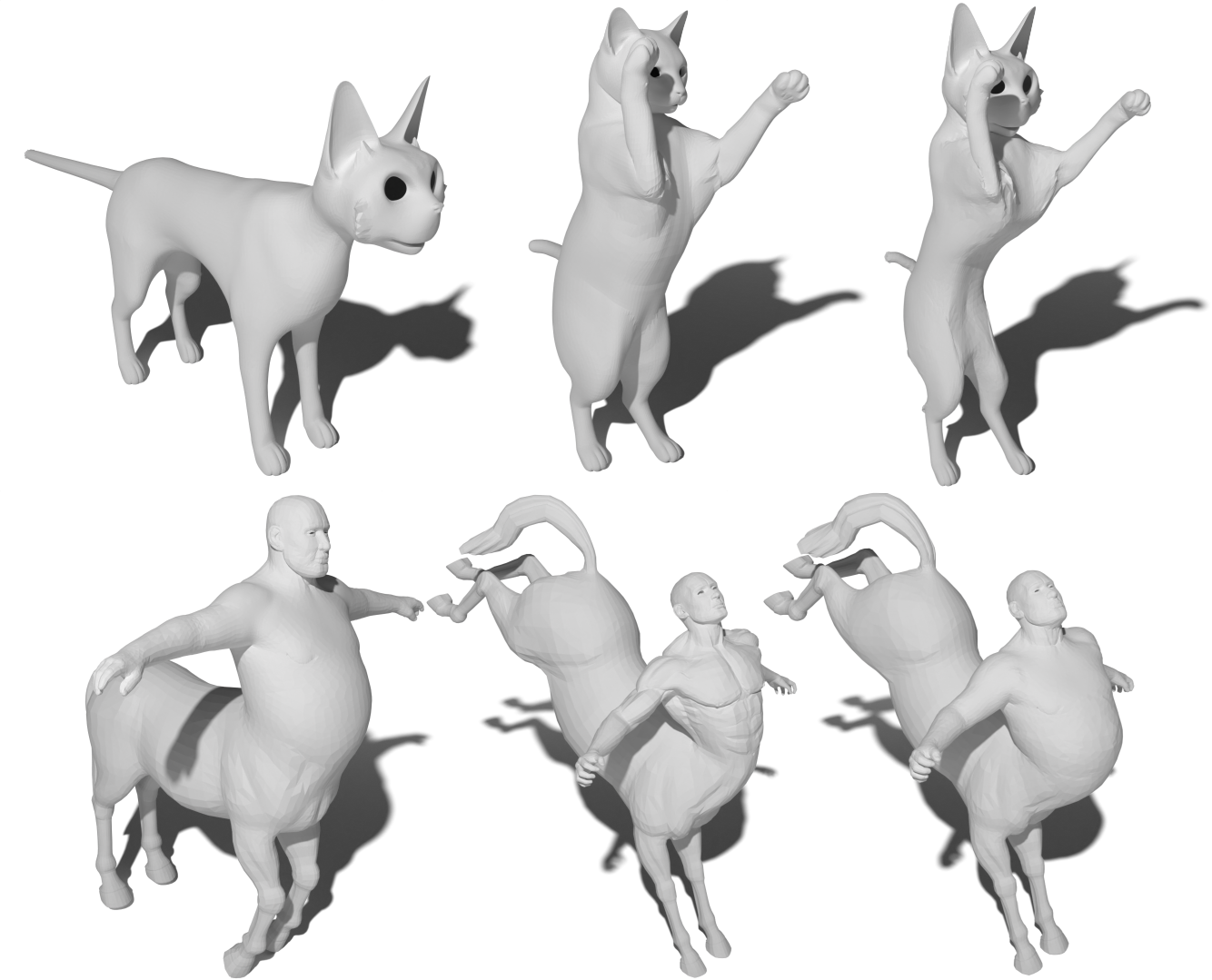}
    \caption{From left to right: a mesh obtained from the rest pose of a TOSCA mesh by manual editing; a mesh from the TOSCA dataset; the manually edited mesh deformed with our method to the pose of the TOSCA mesh.}
    \label{fig:pose-transfer}
\end{figure}

In \Cref{fig:pose-transfer} we show two examples where we deform our edited \texttt{cat} and \texttt{centaur} meshes from the rest pose to another pose defined by a mesh representing the corresponding original subject. Despite some minor visual artifact, we achieve visually plausible results in \textasciitilde 2 seconds of computation for each example. Additionally, as mentioned in \Cref{sec:method}, our method allows us to define the local reference frames onto the triangles, instead of the vertices. By performing the same experiment onto the full-resolution meshes and defining the local reference frames at the vertices (as in ~\cite{melzi:2019:gframes}) we obtained meshes of inferior visual quality, showing the additional benefits in using our pipeline.

\section{Conclusions}\label{sec:conclusions}

We introduced a novel pipeline for efficiently tackling the problem of deforming high-resolution meshes. We exploited a scalable remeshing algorithm to accurately represent a geometry at a lower resolution and improved the existing approach for mapping vertices at high-resolution onto the low-resolution shape. By taking advantage of the correspondence between a mesh and its deformed counterpart, we implemented an efficient local reference frame at triangles using it for representing and deforming the original high-resolution geometry. We validated our approach on multiple sets of data, including a novel dataset we introduced for as-rigid-as-possible deformation tasks. Our extensive set of experiments demonstrates the efficiency and efficacy of our approach in multiple settings, including ARAP deformations, non-rigid isometric transformations, and pose transfer tasks.

In the future, we plan exploring other possible applications of our pipeline, like shape matching and other types of mesh editing tasks. We also intend to explore the possibility of softening the requirements of our method, like the assumption of manifoldness and the graph isomorphism constraint between the low-resolution mesh before and after the deformation. Finally, we mean to probe other approaches for representing the high-resolution geometry with local descriptors, eventually increasing their expressive power and enabling more drastic deformations like topological changes.

\bibliographystyle{eg-alpha-doi}  
\bibliography{egbibsample}        



\appendix

\section{DeFAUST dataset}\label{sec:defaust}

To the best of our knowledge, no dataset exists for quantitative evaluations on deformation methods. For a more exhaustive validation of our method, we introduce the \textbf{De}forming \textbf{FAUST} (DeFAUST) dataset, which we derive from the well-known FAUST dataset for shape matching~\cite{bogo:2014:faust}. The FAUST dataset contains 3D scans of real humans in various poses. Among these shapes, 10 subjects in 10 different poses (including a rest pose for each subject) are registered to a template with about 7000 vertices, meaning that all the 100 meshes are in one-to-one correspondence and share the same connectivity.

\begin{figure}[t]
    \centering
    \includegraphics[width=\linewidth]{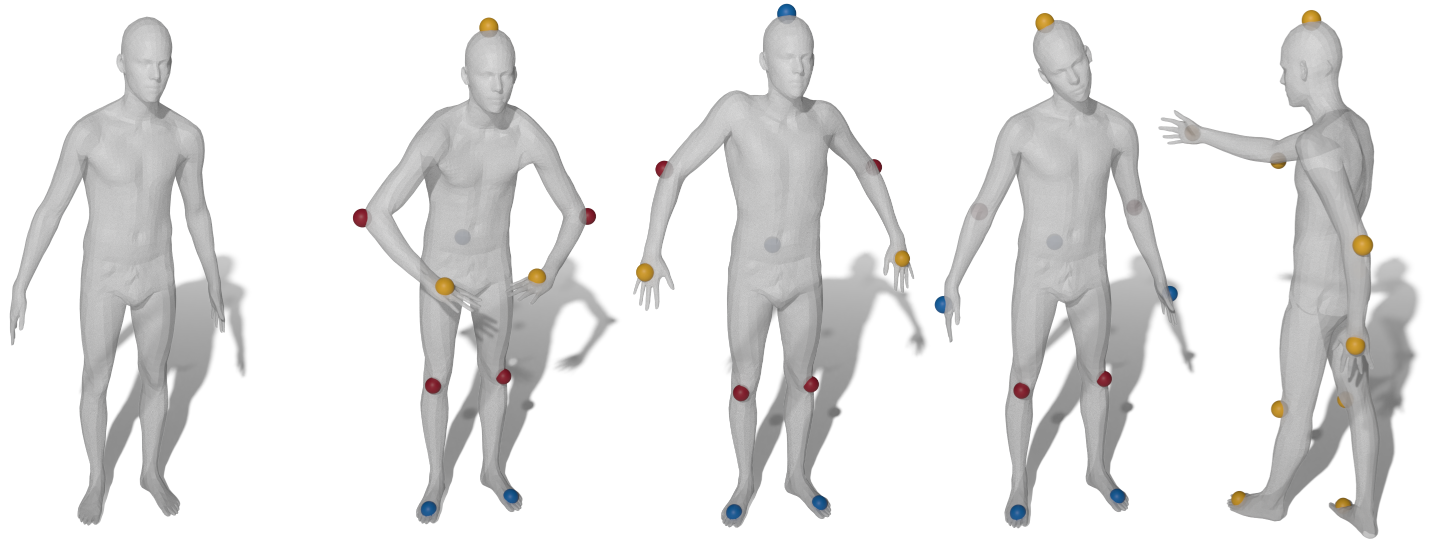}
    \caption{from left to right: the rest pose used in the dataset as the starting position for the deformation; the four target poses used for handle extarction. The landmarks used for the handle generation are color-coded: blue for static handles; yellow for dynamic handles; red for unused handles.}
    \label{fig:faustarap-selection}
\end{figure}

For each subject, we select the rest pose as the starting pose for the deformation and 4 target poses from which we extract the deformation handles. We also handpick some landmarks which we use for driving the handle extraction. \Cref{fig:faustarap-selection} depicts the rest pose and the 4 selected poses, as well as the landmarks that we use for the deformation. Starting from each landmark $\ell$ on the rest pose $\M$, we select 30 random vertices in a neighborhood of $\ell$, obtaining the set $H_\ell$ of handles that drive the deformation of the region associated with the landmark. Then, by using the provided correspondence $\pi : \M \to \M'$ between $\M$ and another target pose $\M'$, we select the handles $H_\ell' = \{ \pi(v) : v \in H_\ell \}$ on $\M'$. We compute the difference
\begin{equation}
    \Delta_\ell
    =
    \sum_{v \in H_\ell} \| v - \pi(v) \|\,,
\end{equation}
and if $\Delta_\ell$ is smaller than some threshold, we mark the positions of the handles $H_\ell$ as static. Otherwise, we mark them as dynamic handles, using the corresponding positions in $H_\ell'$ as their target positions. For the last pose on the right of \Cref{fig:faustarap-selection} we use more landmarks. Differently from the other poses, this pose is not aligned in 3D space with the rest pose, and more information could be needed to solve for a meaningful deformation.

As a result, we obtain a dataset containing 10 different human shapes and 4 sets of deformation handles for each shape. The deformation handles are provided as 3D positions instead of vertex indices, so that they can be used with remeshed and subdivided shapes. Additionally, since we want our dataset to be used also for the evaluation of deformation algorithms on high-resolution meshes, we produce an high-resolution version of the 10 rest pose shapes using the LS3 Loop algorithm~\cite{boye:2010:ls3l} to obtain synthetic shapes with an high vertex count. As noted by Yang \etal~\cite{yang:2021:nfgp}, most deformation algorithms (especially those designed for as-rigid-as-possible deformation tasks) do not perform well on grid-like triangulations like Marching Cubes meshes or shapes obtained by means of regular subdivisions. For this reason, we apply a step of isotropic remeshing as in \cite{hoppe:1993:meshopt} to the subdivided surfaces, obtaining meshes with a vertex count in the range [150k, 200k].

\end{document}